\documentclass[showpacs,preprintnumbers,amsmath,amssymb,aps,prd,nofootinbib,unicode-math,tikz,border=5pt]{revtex4-2}
\usepackage{graphicx, tikz}
\usepackage{amsmath,amssymb,empheq}
\usepackage{epstopdf}
\usepackage{geometry}
\graphicspath{{./Figures/}}

\usepackage{color}%

\usepackage[colorlinks,urlcolor=blue,citecolor=blue]{hyperref}
 
\begin{document}
\usetikzlibrary{arrows.meta}

\title{Weak-Strong Resurgence Duality}

\author{Gerald V. Dunne and Ayssar I. Farah}
\affiliation{Physics Department, University  of Connecticut, Storrs, CT 06269}

%\date{}

\begin{abstract}
We show that there is an explicit resurgent duality between weak and strong coupling expansions when one of the expansions has zero radius of convergence and the other has infinite radius of convergence. This complements the situation where the convergent expansion has finite radius of convergence,  or when both expansions have zero radius of convergence. We illustrate this phenomenon for the Airy and Pearcey catastrophe integrals, and we apply it to two physical examples: the weak and strong coupling expansions of Dyson-Schwinger equations in zero-dimensional scalar $\phi^4$ theories, and the short and long time expansions of the heat kernel trace for the fluctuation operator of the kink-antikink crystal saddle configuration in the Gross-Neveu model.
\end{abstract}

\maketitle

%\newpage

\tableofcontents

\newpage

\section{Introduction}

Resurgent asymptotics is a deep mathematical framework with many applications in physics, such as optics, dynamical systems, fluid dynamics, quantum mechanics, matrix models, string theory and quantum field theory \cite{ecalle3,sauzin,h2plus,alvarez,Zinn-Justin:2004vcw,Zinn-Justin:2004qzw,Marino:2007te,Marino:2012zq,annrev,aniceto}. Its primary advantage is to extract physical information from perturbative expansions which are generically formal factorially divergent series. Resurgence also reveals precise connections between expansions in different parametric regimes \cite{berry-howls}, which furthermore enables significant improvements in continuation methods \cite{cd}. In many applications, one of the strong-coupling or weak-coupling expansions has zero radius of convergence while the other has a finite radius of convergence \cite{leguillou}. There is also a class of physical problems in which both the strong-coupling and weak-coupling expansions have zero radius of convergence \cite{Gukov:2016njj,Costin:2023kla,Adams:2025aad}. In this latter case, modularity plays an important role \cite{Cheng:2018vpl,Cheng:2024vou,Fantini:2024ihf}. Here we discuss resurgent dualities in the situation where one of the weak-coupling or strong-coupling expansions has zero radius of convergence while the other has infinite radius of convergence.  

We first illustrate the idea with the familiar Airy function, the lowest catastrophe integral (the ``fold") and also the next catastrophe integral, the Pearcey integral (the ``cusp"). The short distance expansion has infinite radius of convergence, which can be  written in a particular form that reveals a direct  connection with the expansions at infinity. This behavior extends to higher order generalizations of the Pearcey integral. 
We  apply these results to two physical applications in quantum field theory: (i) a zero-dimensional scalar $\phi^4$ quantum field theory \cite{bender-nonunique, Bender:2023ttu}, for which the path integral generating function $Z[j]$, for coupling to a  current $j$, is directly related to the full Pearcey integral $P(x, y)$ and (ii)  the semiclassical analysis of fluctuations about crystalline solutions to the gap equation of the 2 dimensional Gross-Neveu model \cite{Gross:1974jv,Thies:2006ti,bdt,Dunne:2009zz}. The duality exists between the large order asymptotics of the coefficients of each of the weak-coupling and strong-coupling expansions, and also between the two expansions themselves.
This is an explicit realization of the global analytic continuation idea in \cite{costin-xia}.

\section{Airy Function}
\label{sec:airy}

The Airy function is a simple paradigm of resurgent asymptotics, being the  physical example in which  the Stokes phenomenon was first identified \cite{Airy:1838:ILN,stokes1,stokes2,berry-airy}.
The Airy function solves the second order linear differential equation
\begin{equation}
\frac{d^2 y}{dx^2}-xy=0
\end{equation} and has an integral representation: 
\begin{equation}
\text{Ai}(x) = \frac{1}{\pi} \int^{\infty}_0 dt \cos \left (\frac{t^3}{3}+xt \right).
\end{equation} 
The Airy function has wide applications in physics, in particular optics where it describes caustics and in quantum mechanics where it is characterizes the local behavior of the wavefunction near an isolated turning point. 

\subsection{Airy: large-order/low-order duality as $x\to +\infty$}
\label{sec:airy-strong}

As $x\to +\infty$, the Airy equation, $
y^{\prime\prime}(x)=x\, y(x)$, has two independent solutions with factorially divergent asymptotic series:
 \begin{eqnarray}
y_\pm (x)= \left\{ \begin{matrix} {\rm Ai}(x) \cr  \frac{1}{2}\, {\rm Bi}(x) \end{matrix}  \right\} \sim \frac{e^{\mp \frac{2}{3} x^{\frac{3}{2}}} }{2\, \pi^{\frac{1}{2}}\, x^{\frac{1}{4}} } \sum_{k=0}^\infty \left(\mp1\right)^k \frac{ \Gamma\left(k+\frac{1}{6}\right) \Gamma\left(k+\frac{5}{6}\right)}{(2\pi)\, k!\, \left(\frac{4}{3}\, x^{3/2}\right)^k}
 \label{eq:airy-yplus-yminus}
 \end{eqnarray}
 The 1/2 factor with ${\rm Bi}$ is a normalization convention. 
The expansion coefficients are rational numbers:
\begin{eqnarray}
c_k^{(\pm)}:=\left(\mp1\right)^k \frac{\Gamma\left(k+\frac{1}{6}\right) \Gamma\left(k+\frac{5}{6}\right)}{ 2\pi\, k!\, \left(\frac{4}{3}\right)^k}= \left\{1,\mp \frac{5}{48},\frac{385}{4608}, \mp \frac{85085}{663552}, \dots\right\}
\label{eq:airy-ck}
\end{eqnarray}
The coefficients $c_k^{(\pm)}$ grow factorially in magnitude, and illustrate a generic feature of resurgent functions: they have {\it large-order/low-order resurgence relations},  a duality between the large $k$ behavior of the expansion coefficients $c_k^{(\mp)}$, and the large $x$ expansion of the functions $y_\pm(x)$ \cite{berry-howls,dunne-cern}. 

For example, at large order $(k\to +\infty)$: 
\begin{eqnarray}
c_k^{(-)}\sim  \frac{(k-1)!}{(2\pi) \left(\frac{4}{3}\right)^k}\left(
1- \frac{5}{36} \frac{1}{k}+
\frac{25}{2592}\frac{1}{k^2} + \frac{775}{279936}\frac{1}{k^3}-\dots\right) 
\qquad, \qquad k\to +\infty
\label{eq:airy-ck-large}
\end{eqnarray}
Therefore the large $x$ expansion has zero radius of convergence. This large-order behavior can be re-written in terms of decreasing factorials 
\begin{align}
c_k^{(-)} &\sim \frac{(k-1)!}{(2\pi)\left(\frac{4}{3}\right)^k}
\left(
1-\left(\frac{4}{3}\right)\frac{5}{48}\frac{1}{(k-1)}
+\left(\frac{4}{3}\right)^2\frac{385}{4608}\frac{1}{(k-1)(k-2)}
\right. \notag\\
&\qquad\left.
-\left(\frac{4}{3}\right)^3\frac{85085}{663552}\frac{1}{(k-1)(k-2)(k-3)}
+\dots
\right)
\qquad, \qquad k\to\infty .
\label{eq:ck-resurgent}
\end{align}
When written in this form, the coefficients, $\{1, -5/48, 385/4608, -85085/663552, ...\}$, of the subleading terms of the  large $k$ behavior in \eqref{eq:ck-resurgent}, match the coefficients of the subleading terms in  \eqref{eq:airy-ck} of the large $x$ behavior of the other saddle solution $y_+(x)$. The factor $\frac{4}{3}$  in \eqref{eq:ck-resurgent}  corresponds to the difference between the ``actions'', $\pm \frac{2}{3}$, of the two Airy saddle points. This is a generic feature of resurgent functions: the asymptotic expansion about one saddle point is directly related to the expansion about other saddle points \cite{berry-howls}.

\subsection{Airy: large-order duality as $x\to 0$}
\label{sec:airy-weak}

The behavior of the Airy function is completely different as $x\to 0$. The $x\to 0$ expansion has {\it infinite} radius of convergence:
\begin{eqnarray}
    {\rm Ai}(x)=\frac{1}{3^{2/3} \pi}\sum_{n=0}^\infty \sin\left(\frac{2\pi(n+1)}{3}\right)
    \frac{\Gamma\left(\frac{n+1}{3}\right)}{n!} (3^{1/3}x)^{n}
    \qquad, \qquad x\to 0
    \label{eq:airy-small-x}
\end{eqnarray}
The large-order, $n\to \infty$, behavior of these small $x$ expansion coefficients is
\begin{eqnarray}
    \frac{\Gamma\left(\frac{n+1}{3}\right)}{n!}
    &\sim & \frac{\sqrt{\pi } 2^{\frac{2 n}{3}+\frac{7}{6}} 3^{-n-\frac{1}{2}}}{\Gamma \left(\frac{2
   n}{3}+\frac{7}{6}\right)}
   \left[1+\frac{5}{48 n} -\frac{455}{4608 n^2} +\frac{19565}{663552 n^3}+\frac{12350065}{127401984 n^4}+\dots \right] \quad, \quad n\to\infty \\
    &\sim & \frac{\sqrt{\pi } 2^{\frac{2 n}{3}+\frac{7}{6}} 3^{-n-\frac{1}{2}}}{\Gamma \left(\frac{2
   n}{3}+\frac{7}{6}\right)}
   \left[1+\left(\frac{2}{3}\right) \frac{5}{48 \left(\frac{2 n}{3}+\frac{7}{6}\right)}+\left(\frac{2}{3}\right)^2 \frac{385}{4608
   \left(\frac{2 n}{3}+\frac{7}{6}\right) \left(\frac{2 n}{3}+\frac{13}{6}\right)}
   \right.
   \nonumber\\
   && \left. 
   +\left(\frac{2}{3}\right)^3 \frac{85085}{663552
   \left(\frac{2 n}{3}+\frac{7}{6}\right) \left(\frac{2 n}{3}+\frac{13}{6}\right) \left(\frac{2
   n}{3}+\frac{19}{6}\right)}+\dots \right]
   \label{eq:airy-large-n}
\end{eqnarray}
We see that the coefficients of the subleading corrections in \eqref{eq:airy-large-n} match the subleading corrections \eqref{eq:airy-ck-large} of the large-order growth of the large $x$ expansion coefficients, as well as the large $x$ expansion coefficients themselves in \eqref{eq:airy-ck}. Thus, there is a direct connection between the coefficients of the small and large $x$ expansions, which have infinite and zero radius of convergence, respectively.

\subsection{Refined duality between Airy large $x$ and small $x$ expansions}

Another way to see this duality between the small $x$ and large $x$ expansions is to note that the $x\to 0$ expansion naturally splits into two parts:
\begin{eqnarray}
    {\rm Ai}(x)=
    {\rm Ai}(0)\sum_{n=0}^\infty x^{3n} \frac{3^n \Gamma\left(n+\frac{1}{3}\right)}{\Gamma(3n+1) \, \Gamma\left(\frac{1}{3}\right)}
    +{\rm Ai}^\prime(0)\sum_{n=0}^\infty x^{3n+1} \frac{3^n \Gamma\left(n+\frac{2}{3}\right)}{\Gamma(3n+2) \, \Gamma\left(\frac{2}{3}\right)}
    \label{eq:airy-zero}
\end{eqnarray}
where ${\rm Ai}(0)=1/(3^{2/3} \Gamma(2/3))$, and ${\rm Ai}^\prime(0)=-1/(3^{1/3} \Gamma(1/3))$. The expansion coefficients in \eqref{eq:airy-zero} {\it decrease} factorially as $n\to +\infty$. This can be written in the following form, as a series of inverses of increasing factorials:
\begin{eqnarray}
    b_n^{(+)}& := & {\rm Ai}(0)\frac{3^n \Gamma\left(n+\frac{1}{3}\right)}{\Gamma(3n+1) \, \Gamma\left(\frac{1}{3}\right)} 
    \nonumber\\
    &=& \frac{1}{\sqrt{3\pi}} \frac{\left(\frac{2}{3}\right)^{2n+1/6} }{\Gamma\left(2n+\frac{7}{6}\right)}\left[1+\frac{\left(\frac{2}{3}\right) \frac{5}{48}}{\left(2n+\frac{7}{6}\right)} +\frac{\left(\frac{2}{3}\right)^2 \frac{385}{4608}}{\left(2n+\frac{7}{6}\right)\left(2n+\frac{13}{6}\right)} 
    +  \frac{\left(\frac{2}{3}\right)^3 \frac{85085}{663552}}{\left(2 n+\frac{7}{6}\right) \left(2 n+\frac{13}{6}\right) \left(2 n+\frac{19}{6}\right)}+\dots \right]
    \\
    b_n^{(-)}& := & {\rm Ai}^\prime(0)\frac{3^n \Gamma\left(n+\frac{2}{3}\right)}{\Gamma(3n+2) \, \Gamma\left(\frac{2}{3}\right)} 
    \nonumber\\
    &=& -\frac{1}{\sqrt{3\pi}}\frac{\left(\frac{2}{3}\right)^{2n+5/6} }{\Gamma\left(2n+\frac{11}{6}\right)}\left[1+ \frac{\left(\frac{2}{3}\right) \frac{5}{48}}{\left(2n+\frac{11}{6}\right)} + \frac{\left(\frac{2}{3}\right)^2 \frac{385}{4608}}{\left(2n+\frac{11}{6}\right)\left(2n+\frac{17}{6}\right)} +  \frac{\left(\frac{2}{3}\right)^3 \frac{85085}{663552}}{\left(2 n+\frac{11}{6}\right) \left(2 n+\frac{17}{6}\right) \left(2 n+\frac{23}{6}\right)}+\dots \right]
    \label{eq:airy-zero-large}
\end{eqnarray}
Thus, for each set of coefficients in the refined small $x$ expansion \eqref{eq:airy-zero}, we observe
that the subleading corrections to the large order behavior involve exactly the same rational coefficients that appear in the $x\to +\infty$ expansion \eqref{eq:airy-ck} and in the $n\to \infty$ corrections \eqref{eq:ck-resurgent}.

This can be considered a generalization of Darboux's fundamental result that the expansion about a point encodes information about the expansion at the neighboring singularities \cite{henrici}: here, the nearest singularity for the expansion about $x=0$ is at infinity.

\section{Bessel Function}
\label{sec:bessel}

The correspondence between the $x\to +\infty$ and $x\to 0$ expansions of the Airy function described in the previous  section is a special case of a more general result for Bessel functions (the Airy functions can be expressed in terms of Bessel functions). 

\subsection{Bessel: large-order/low-order duality as $x\to +\infty$}
\label{sec:bessel-strong}

As $x\to +\infty$, the modified Bessel function $I_\nu(x)$ grows  exponentially, with a square-root prefactor, multiplying a factorially divergent Laurent series:
\begin{eqnarray}
    I_\nu(x)\sim \frac{e^x}{\sqrt{2\pi x}} \sum_{k=0}^\infty a_k(\nu) \frac{1}{x^k} \qquad, \qquad x\to +\infty
    \label{eq:bessel-large}
\end{eqnarray}
The expansion coefficient $a_k(\nu)$ is a $k^{\rm th}$ order polynomial in the square of the Bessel index $\nu$:
\begin{eqnarray}
    a_k(\nu)&=&
    \frac{\cos(\pi \nu)}{\pi} \frac{\Gamma\left(k+\frac{1}{2}-\nu\right)\Gamma\left(k+\frac{1}{2}+\nu\right)}{2^k \Gamma(k+1)}
    \\
    &=&\left\{1, -\frac{(4\nu^2-1)}{8}, \frac{(4\nu^2-1^2)(4\nu^2-3^2)}{128}, 
    -\frac{(4\nu^2-1^2)(4\nu^2-3^2)(4\nu^2-5^2)}{3072}, \dots\right\}
    \label{eq:bessel-an}
\end{eqnarray}
At large order, $k\to +\infty$, for a generic $\nu$, the coefficients $a_k(\nu)$ grow factorially in magnitude.
\begin{eqnarray}
    a_k(\nu)
    &\sim& \frac{\cos(\pi \nu)}{\pi}\frac{\Gamma(k)}{ 2^k}\left(1+\frac{2(4\nu^2-1)}{8\,(k-1)}+\frac{2^2 (4\nu^2-1^2)(4\nu^2-3^2)}{128\,(k-1)(k-2)}+ \frac{2^3 (4\nu^2-1^2)(4\nu^2-3^2)(4\nu^2-5^2)}{3072\,(k-1)(k-2)(k-3)}\dots\right)
    \label{eq:bessel-ak}
\end{eqnarray}
Note that the coefficients of the subleading power-law terms in the large order behavior in \eqref{eq:bessel-ak} are equal to the low order coefficients in \eqref{eq:bessel-an}, up to a factor of $(-2)^k$ where 2 is the difference between the coefficients $\pm 1$ in the exponent of $e^{\pm x}$. This is another example of the generic large-order/low-order resurgence relation \cite{berry-howls,dunne-cern}.

\subsection{Bessel: large-order duality as $x\to 0$}
\label{sec:bessel-weak}

As $x\to 0$, the Bessel function $I_\nu(x)$ has an expansion with {\it infinite} radius of convergence:
\begin{eqnarray}
    I_\nu(x)=\left(\frac{x}{2}\right)^\nu \sum_{n=0}^\infty \frac{x^{2n}}{4^n\, \Gamma(n+1) \,\Gamma(n+\nu+1)}
    \label{eq:bessel-zero}
\end{eqnarray}
The expansion coefficients {\it decrease} factorially with the expansion order, and this may be written as an expansion in the inverses of increasing factorials:
\begin{eqnarray}
    \frac{1}{\Gamma(n+1)\Gamma(n+\nu+1)}&\sim & \frac{2^{2n+2\nu+1}}{\sqrt{\pi}\, \Gamma\left(2n+\nu+\frac{3}{2}\right)}
    \left[1-\frac{(4\nu^2-1)}{8\left(2n+\nu+\frac{3}{2}\right)} +\frac{(4\nu^2-1)(4\nu^2-9)}{128\left(2n+\nu+\frac{3}{2}\right)\left(2n+\nu+\frac{5}{2}\right)} \right.
    \nonumber\\
    && \hskip 2cm  \left. 
    -\frac{(4\nu^2-1)(4\nu^2-9)(4\nu^2-25)}{3072\left(2n+\nu+\frac{3}{2}\right)\left(2n+\nu+\frac{5}{2}\right)\left(2n+\nu+\frac{7}{2}\right)}+\dots
    \right]
\end{eqnarray}
Once again, we see that the coefficients of the subleading corrections to the large-order growth of the factorially decreasing coefficients of the small $x$ expansion match the coefficients of the subleading corrections to the large-order growth of the factorially growing coefficients of the large $x$ expansion.

\section{Pearcey Function}
\label{sec:pearcey}
The Pearcey integral \cite{paris} is the canonical diffraction integral for a cusp catastrophe, typically written as
\begin{equation}
P(x,y)=\int_{-\infty}^{\infty}\exp\big(i(\phi^{4}+x \phi^{2}+y \phi )\big)\,dt,
\label{eq:pearcey-full}
\end{equation}
whose stationary-phase geometry produces the characteristic cusp caustic (\href{https://dlmf.nist.gov/36.2.E14}{dlmf/36.2.14}). The asymptotics of $P(x,y)$ is  governed by multiple coalescing saddle points and rich Stokes phenomena \cite{paris}. 
We first consider two special cases of the Pearcey integral.

\subsection{First special Pearcey function: $P_1$}
\label{sec:p1}

In this section we analyze the following 1-variable case of the Pearcey integral
\begin{eqnarray}
    P_1(y):=\int_0^\infty d\phi\exp\left[-\phi^4-y\, \phi\right]
    \label{eq:pearcey-j}
\end{eqnarray}
This integral has been studied in the context of Schwinger-Dyson equations for zero-dimensional massless $\phi^4$ quantum field theory, with $y$ playing the role of the rescaled  current \cite{bender-nonunique,Bender:2023ttu}. See also section \ref{sec:dse}.

\subsubsection{Pearcey 1: large-order behavior as $y\to +\infty$}
\label{sec:pearcey1-strong}

The formal $y\to+\infty$ expansion of $P_1(y)$ 
leads to the factorially divergent asymptotic expansion:
\begin{eqnarray}
    P_1(y)\sim \sum_{k=0}^\infty(-1)^k \frac{\Gamma(4k+1)}{k!} \frac{1}{y^{4k+1}} 
    \qquad, \qquad y\to +\infty
    \label{eq:p1-strong}
\end{eqnarray}
The large order $(k\to+\infty)$ behavior of these coefficients can be expressed as a series of decreasing factorials:
\begin{eqnarray}
    \frac{\Gamma(4k+1)}{k!} &\sim& 
     \sqrt{\frac{2}{\pi}} \Gamma\left(3k+\frac{1}{2}\right)\left(\frac{4^4}{3^3}\right)^k
    \left(1-\frac{7}{144 k}+\frac{49}{41472 k^2}+\frac{47033}{17915904 k^3}-\dots\right)
     \quad, \quad k\to +\infty \\
   &\sim& 
    \sqrt{\frac{2}{\pi}} \Gamma\left(3k+\frac{1}{2}\right)\left(\frac{4^4}{3^3}\right)^k\left[1-\frac{7}{48}\frac{1}{\left(3k-\frac{1}{2}\right)} +\frac{385}{4608}\frac{1}{\left(3k-\frac{1}{2}\right)\left(3k-\frac{3}{2}\right)}
    \right.
    \nonumber\\
    && \hskip 4cm \left. 
    -\frac{39655}{663552}\frac{1}{\left(3k-\frac{1}{2}\right)\left(3k-\frac{3}{2}\right)\left(3k-\frac{5}{2}\right)}+\dots \right]
     \quad, \quad k\to +\infty
    \label{eq:p1-large-k}
\end{eqnarray}

\subsubsection{Pearcey 1: large-order duality as $y\to 0$}
\label{sec:pearcey1-weak}

The  $y\to 0$ expansion of $P_1(y)$ is factorially convergent, with an infinite radius of convergence:
\begin{eqnarray}
    P_1(y)=\frac{1}{4} \sum_{n=0}^\infty (-1)^n \frac{\Gamma\left(\frac{n+1}{4}\right)}{n!} {y^n} 
    \qquad, \qquad y\to 0
    \label{eq:p1-weak}
\end{eqnarray}
The coefficients in \eqref{eq:p1-weak} can be expressed as a series of inverses of increasing factorials:
\begin{eqnarray}
 \frac{1}{4} \frac{\Gamma\left(\frac{n+1}{4}\right)}{n!} &\sim& 
 \sqrt{\frac{\pi}{2}} 
 \frac{\left(\frac{27}{256}\right)^{(n+1)/4}}{\Gamma\left(\frac{3(n+1)}{4}+\frac{1}{2}\right)}
 \left(1+\frac{7}{36 n}-\frac{455}{2592 n^2}-\frac{3185}{279936 n^3}+\frac{13442065}{40310784
   n^4}+\dots \right)
   \quad, \quad n\to\infty
   \\
   &\sim&
 \sqrt{\frac{\pi}{2}} 
 \frac{\left(\frac{27}{256}\right)^{(n+1)/4}}{\Gamma\left(\frac{3(n+1)}{4}+\frac{1}{2}\right)}\left[1+\frac{7}{48}\frac{1}{\left(\frac{3(n+1)}{4}+\frac{1}{2}\right)}
 +\frac{385}{4608}\frac{1}{\left(\frac{3(n+1)}{4}+\frac{1}{2}\right)\left(\frac{3(n+1)}{4}+\frac{3}{2}\right)}
    \right.
    \nonumber\\
    && \hskip 2cm \left. 
    +\frac{39655}{663552}\frac{1}{\left(\frac{3(n+1)}{4}+\frac{1}{2}\right)\left(\frac{3(n+1)}{4}+\frac{3}{2}\right)\left(\frac{3(n+1)}{4}+\frac{5}{2}\right)}+\dots \right]
    \label{eq:p1-weak-growth}
\end{eqnarray}
We see that when the subleading corrections to the leading factorial decay of the coefficients of the small $y$ expansion of $P_1(y)$ are expressed in terms of increasing factorials, as in \eqref{eq:p1-weak-growth}, the coefficients of the subleading corrections match the coefficients of the  subleading corrections to the large order growth of the coefficients of the large $y$ expansion of $P_1(y)$ in \eqref{eq:p1-large-k}. 

This means that the small and large $y$ expansions of $P_1(y)$ are related. In fact, the small and large $y$ expansions map into one another under the replacement:
\begin{eqnarray}
    k\longleftrightarrow -\left(\frac{n+1}{4}\right)
    \label{eq:p1-map}
\end{eqnarray}
This mapping is explained below in section \ref{sec:mellin} using the Mellin transform approach to asymptotics.

\subsubsection{Refined duality for Pearcey 1}

Another perspective of the resurgent duality between the large order behavior of the large $y$ and small $y$ expansions is given by noting that 
$P_1(y)$ can be expressed as the sum of 4 generalized hypergeometric functions:
\begin{eqnarray}
P_1(y)&=&\Gamma
   \left(\frac{5}{4}\right) \,
   _0F_2\left(;\frac{1}{2},\frac{3}{4};\frac{y^4}{256}\right)-\frac{1}{4} y \sqrt{\pi }  \,
   _0F_2\left(;\frac{3}{4},\frac{5}{4};\frac{y^4}{256}\right)+ 
   \nonumber\\
   && + \frac{1}{8} y^2 \Gamma
   \left(\frac{3}{4}\right) \,
   _0F_2\left(;\frac{5}{4},\frac{3}{2};\frac{y^4}{256}\right)-\frac{1}{24} y^3 \,
   _1F_3\left(1;\frac{5}{4},\frac{3}{2},\frac{7}{4};\frac{y^4}{256}\right)
   \label{eq:p1-hyper}
\end{eqnarray}
Note that $P_1(y)$ satisfies a fourth order ordinary differential equation. Each of the four generalized hypergeometric functions in \eqref{eq:p1-hyper} has an asymptotic expansion as $y\to +\infty$:
\begin{eqnarray}
   \Gamma\left(\frac{5}{4}\right)\, _0F_2\left(;\frac{1}{2},\frac{3}{4};\frac{y^4}{256}\right) 
   &\sim& 
   \frac{\sqrt{\pi/24}}{(2y)^{1/3}}\exp\left(3 \left(\frac{y}{4}\right)^{4/3}\right)\left(1+\frac{7/48}{\left(3 \left(\frac{y}{4}\right)^{4/3}\right)}
   +\frac{385/4608}{\left(3 \left(\frac{y}{4}\right)^{4/3}\right)^2}+\dots \right)
   \nonumber\\
   \frac{\sqrt{\pi}}{4}\,y \, _0F_2\left(;\frac{3}{4},\frac{5}{4};\frac{y^4}{256}\right) &\sim & \frac{\sqrt{\pi/24}}{(2y)^{1/3}} 
\exp\left(3 \left(\frac{y}{4}\right)^{4/3}\right)\left(1+\frac{7/48}{\left(3 \left(\frac{y}{4}\right)^{4/3}\right)}
   +\frac{385/4608}{\left(3 \left(\frac{y}{4}\right)^{4/3}\right)^2}+\dots \right)
  \nonumber \\
  \frac{\Gamma\left(\frac{3}{4}\right)}{8}\, y^2 \, _0F_2\left(;\frac{5}{4},\frac{3}{2};\frac{y^4}{256}\right) &\sim& \frac{\sqrt{\pi/24}}{(2y)^{1/3}}  \exp\left(3 \left(\frac{y}{4}\right)^{4/3}\right)\left(1+\frac{7/48}{\left(3 \left(\frac{y}{4}\right)^{4/3}\right)}
   +\frac{385/4608}{\left(3 \left(\frac{y}{4}\right)^{4/3}\right)^2}+\dots \right)
   \nonumber \\
   \frac{1}{24} y^3 \,
   _1F_3\left(1;\frac{5}{4},\frac{3}{2},\frac{7}{4};\frac{y^4}{256}\right) &\sim & 
   \frac{\sqrt{\pi/24}}{(2y)^{1/3}}  \exp\left(3 \left(\frac{y}{4}\right)^{4/3}\right)\left(1+\frac{7/48}{\left(3 \left(\frac{y}{4}\right)^{4/3}\right)}
   +\frac{385/4608}{\left(3 \left(\frac{y}{4}\right)^{4/3}\right)^2}+\dots \right)
   \nonumber\\
   &&
-\left(\frac{1}{y}-\frac{24}{y^5}+\frac{20160}{y^9}-\frac{79833600}{y^{13}}+\frac{871782912000}{y^{17}}+\dots\right)
   \label{eq:p1-asymptotics}
\end{eqnarray}
Observe that in \eqref{eq:p1-asymptotics} the exponential piece is the same, for all four terms. Therefore, when combined with the signs in \eqref{eq:p1-hyper} these exponential terms all cancel, leaving just a pure 
power series term which matches the large $y$ expansion in \eqref{eq:p1-strong}.
However, we also observe that the coefficients of each individual exponential term in \eqref{eq:p1-asymptotics} match the coefficients of the subleading terms in the factorial large order growth of the coefficients of the power series term, as in \eqref{eq:p1-large-k}. In addition, these same coefficients match the coefficients of the subleading terms in the {\it inverse factorial} large order growth of the coefficients of the small $y$ power series term, as shown in \eqref{eq:p1-weak-growth}. Thus, in addition to the usual large-order/low-order duality \cite{berry-howls}, there is also an explicit relation between the large-order asymptotics of the large $y$ and small $y$ expansions.

\subsection{Second special Pearcey function: $P_2$}
\label{sec:p2}

A second special case of the Pearcey integral is the following:
\begin{align} \label{eqn:Pearcey2}
P_2(x)&=\int_0^\infty d\phi\, \exp\left[-\phi^4-x\, \phi^2\right] \\
&= \frac{1}{4} e^{x^2/8} \sqrt{x}\, K_{\frac{1}{4}}(x^2/8)  \qquad,\qquad x>0 
\end{align}

\subsubsection{Pearcey 2: large-order/low-order duality as $x\to +\infty$}
\label{sec:pearcey2-strong}

The  asymptotic expansion of $P_2(x)$ as  $x\to+\infty$ is:
\begin{eqnarray}
P_2(x)&\sim &
\frac{1}{2} \sum_{k=0}^\infty (-1)^k \frac{\Gamma\left(2k+\frac{1}{2}\right)}{k!} \frac{1}{x^{2k+1/2}}
\quad,\quad x\to+\infty
\\
&\sim &\frac{1}{2}\sqrt{\frac{\pi}{x}}\left(1-\frac{3}{16(x/2)^2}+\frac{105}{512 (x/2)^4}-\frac{3465}{8192 (x/2)^6}+ \dots\right) 
\label{eq:p2-large-x}
\end{eqnarray}
The large order ($k\to +\infty$) growth of these coefficients is:
\begin{eqnarray}
    \frac{\Gamma\left(2k+\frac{1}{2}\right)}{k!}  &\sim&
    \frac{1}{\sqrt{2\pi}}\, 4^k\, \Gamma(k)\left(
    1-\frac{3}{16 k}+\frac{9}{512 k^2}+\frac{39}{8192 k^3}-\frac{549}{524288
   k^4}+\dots \right)
   \quad, \quad k\to +\infty
   \\
   &\sim&
   \frac{1}{\sqrt{2\pi}}\, 4^k\, \Gamma(k)\left(
    1-\frac{3}{16 (k-1)}+\frac{105}{512 (k-1)(k-2)}+\frac{3465}{8192 (k-1)(k-2)(k-3)}-\dots \right)
    \label{eq:p2-large-k}
\end{eqnarray}
We see that when the subleading corrections to the leading factorial growth of the coefficients of the large $x$ expansion of $P_2(x)$ are expressed in terms of decreasing factorials, as in \eqref{eq:p2-large-k}, the coefficients of the subleading corrections match the coefficients of the  large $x$ expansion of $P_2(x)$ in \eqref{eq:p2-large-x}.
This is another example of the generic large-order/low-order resurgence relation \cite{berry-howls,dunne-cern}.

\subsubsection{Pearcey 2: large-order duality as $x\to 0$}
\label{sec:pearcey2-weak}

The $x\to 0^+$ expansion of $P_2(x)$ is factorially convergent, with an infinite radius of convergence:
\begin{eqnarray}
P_2(x)
&=& \frac{1}{4}\sum_{n=0}^\infty (-x)^n \frac{\Gamma(n/2+1/4)}{\Gamma(n+1)} 
\label{eq:p2-small-x}
\end{eqnarray}
The large order behavior of the coefficients is factorially decreasing:
\begin{eqnarray}
\frac{\Gamma(n/2+1/4)}{\Gamma(n+1)}
&=&
\frac{\sqrt{\pi}}{2^n} \frac{1}{\Gamma(n/2+5/4)} \left(1+\frac{3}{8 n}-\frac{15}{128 n^2}-\frac{15}{1024 n^3}+\frac{1515}{32768
   n^4}+\dots \right)
   \quad,\quad n\to+\infty
   \\
&=& \frac{\sqrt{\pi}}{2^n} \frac{1}{\Gamma(n/2+5/4)} \left(1+\frac{3}{16(n/2+5/4)}+\frac{105}{512 (n/2+5/4)(n/2+9/4)} \right.
\nonumber\\
&&\left. \hskip 2cm 
+\frac{3465}{8192 (n/2+5/4)(n/2+9/4)(n/2+13/4)}+\dots \right) 
\quad,\quad n\to+\infty
\label{eq:p2-large-n}
\end{eqnarray}
Once again, we see that when the subleading corrections to the leading factorial decay of the coefficients of the small $x$ expansion of $P_2(x)$ are expressed in terms of increasing factorials, as in \eqref{eq:p2-large-n}, the coefficients of the subleading corrections match the coefficients of the large $x$ expansion of $P_2(x)$ in \eqref{eq:p2-large-x}. 

This means that the small and large $x$ expansions of $P_2(x)$ are related. In fact, the small and large $x$ expansions map into one another under the replacement:
\begin{eqnarray}
    k\longleftrightarrow -\left(\frac{n+1/2}{2}\right)
    \label{eq:p2-map}
\end{eqnarray}
This mapping is explained below in section \ref{sec:mellin} using the Mellin transform approach to asymptotics.

\section{Physical Application: Dyson-Schwinger Equations}
\label{sec:dse}

These results for the asymptotics of the Pearcey function can be used to relate the strong-coupling and weak-coupling expansions of zero-dimensional models of a scalar $\phi^4$ quantum field theory path integral \cite{bender-nonunique,Bender:2023ttu}. Consider the zero-dimensional generating function
\begin{eqnarray}
    Z[j]=N_1 \int^{\infty}_{-\infty} d\phi \ \phi^{2n} \exp \left (-\frac{1}{2}m^2\phi^2 -\frac{1}{4}\lambda \phi^4 +\phi\, j \right)
\end{eqnarray}
with normalization factor
\begin{equation}
    N^{-1}_1 = \int^{\infty}_{-\infty} d\phi\exp \left (-\frac{1}{2}m^2\phi^2 -\frac{1}{4}\lambda \phi^4 \right).
\end{equation}
Physically, the variable $j$ corresponds to a current, and derivatives of $Z[j]$ with respect to $j$ generate the multi-point Green's functions of this zero-dimensional $\phi^4$ path integral. By suitable scaling, $Z[j]$ can be expressed in terms of the full Pearcey function in \eqref{eq:pearcey-full}.
Explicitly, the $2n-$point Green's function is given by \cite{bender-nonunique}: 
\begin{equation}
    G_{2n} = N_1 \int^{\infty}_{-\infty} d\phi \ \phi^{2n} \exp \left (-\frac{1}{2}m^2\phi^2 -\frac{1}{4}\lambda \phi^4 \right)
\end{equation}
Up to a factor of $N_1$, this is exactly what we have in the second special case of the Pearcey integral (\ref{eqn:Pearcey2}), with suitable scaling.
The weak-coupling expansion, $\lambda\ll m^4$, is factorially divergent
\begin{equation}
    G_{2n} = N_1 \int^{\infty}_{-\infty} d\phi \ \phi^{2n}e^{-\frac{1}{2}m^2 \phi^2} \sum^{\infty}_{p=0}\frac{(-\frac{1}{4}\lambda \phi^4)^p}{p!} \sim N_1 \left(\frac{2}{m^2}\right)^{n+1/2} \sum^{\infty}_{p=0} \left(-\frac{\lambda}{m^4}\right)^p \frac{\Gamma(2p + n +1/2)}{\Gamma(p+1)}
    \label{eq:ds-weak}
\end{equation}
The strong-coupling expansion, $\lambda\gg m^4$, is factorially convergent
\begin{equation}
    G_{2n} = N_1 \int^{\infty}_{-\infty} d\phi \ \phi^{2n}e^{-\frac{1}{4}\lambda \phi^4} \sum^{\infty}_{p=0}\frac{(-\frac{1}{2}m^2 \phi^2)^p}{p!} \sim N_1 \frac{1}{2}\left(\frac{4}{\lambda} \right)^{n+1/2} \sum^{\infty}_{p=0} \left(-\frac{m^2}{\lambda^{1/2}} \right)^p \frac{\Gamma \left(\frac{p}{2} + \frac{n}{2} + \frac{1}{4} \right)}{\Gamma(p+1)}
    \label{eq:ds-strong}
\end{equation}

The factorial divergence of the coefficients of the weak coupling expansion \eqref{eq:ds-weak} can be written in the standard resurgent decreasing factorial form. For each $n$ we have the large $p$ behavior:
\begin{eqnarray}
    \frac{\Gamma(2p + n +1/2)}{\Gamma(p+1)} &\sim&
    \frac{2^{2p+n-1/2}\, \Gamma(p+n)}{\sqrt{\pi}}\left[ 1-\frac{(2n-1)(2n-3)}{16p} +
    \frac{(2n+3)(2n-3)(2n+1)(2n-1)}{512p^2}+\dots \right]
    \\
    &\sim&
      \frac{2^{2p+n-1/2}\, \Gamma(p+n)}{\sqrt{\pi}}\left[ 1-\frac{(2n-1)(2n-3)}{16(p+n-1)} +
    \frac{(2n-1)(2n-3)(2n-5)(2n-7)}{512(p+n-1)(p+n-2)}+\dots \right]
    \label{eq:ds-weak-resurgence}
\end{eqnarray}
In the second line we have expressed the factorially divergent large $p$ expansion as a sum of decreasing factorials.

The factorial convergence of the coefficients of the strong coupling expansion \eqref{eq:ds-strong} can be written in terms of inverses of increasing factorials. For each $n$ we have the large $p$ behavior:
\begin{eqnarray}
    \frac{\Gamma\left(\frac{p}{2} + \frac{n}{2} +\frac{1}{4}\right)}{\Gamma(p+1)} &\sim&
    \frac{\sqrt{\pi}\, 2^{-p}}{\Gamma\left(\frac{p}{2} - \frac{n}{2} +\frac{5}{4}\right)}\left[ 1+\frac{(2n-1)(2n-3)}{8p} +
    \frac{(2n+1)(2n-1)(2n-3)(2n-5)}{128p^2}+\dots \right]
    \\
    &\sim&
     \frac{\sqrt{\pi}\, 2^{-p}}{\Gamma\left(\frac{p}{2} - \frac{n}{2} +\frac{5}{4}\right)}\left[ 1+\frac{(2n-1)(2n-3)}{16 \left(\frac{p}{2} + \frac{n}{2} +\frac{5}{4}\right)} +
    \frac{(2n-1)(2n-3)(2n-5)(2n-7)}{512\left(\frac{p}{2} + \frac{n}{2} +\frac{5}{4}\right)\left(\frac{p}{2} + \frac{n}{2} +\frac{9}{4}\right)}+\dots \right]
    \label{eq:ds-strong-resurgence}
\end{eqnarray}
Note the correspondence between the coefficients of the subleading large order corrections of the weak and strong coupling expansions in \eqref{eq:ds-weak-resurgence} and \eqref{eq:ds-strong-resurgence}. 
This illustrates the weak/strong resurgent duality for the full 2-variable Pearcey function. Physically, this implies that the weak and strong coupling expansions of the Green's functions are directly related. 

\section{Physical Application: Fluctuations about Kink-Antikink Crystals in the Gross-Neveu Model}
\label{sec:wlgn}

The Gross-Neveu model \cite{Gross:1974jv} is a 2 dimensional quantum field theory that has important features in common with quantum chromodynamics (QCD): it is asymptotically free, it exhibits chiral symmetry breaking and it has a nontrivial finite temperature and density phase diagram \cite{Thies:2006ti,bdt}. In the semiclassical analysis of crystalline solutions to the Gross-Neveu gap equation, one encounters the heat kernel trace of the operator describing fluctuations about the classical kink-antikink crystal \cite{Dunne:2009zz}. For a certain choice of crystal period, this heat kernel trace is described by the function:
\begin{eqnarray}
    \mathcal K(t)=\sqrt{2\pi t}\, I_{1/4}(t)\,  I_{-1/4}(t)
    \label{eq:wlgn-bessel}
\end{eqnarray}
where $I_{\pm 1/4}(t)$ are modified Bessel functions. In the QFT context one is interested in the analytic continuation of this function from small to large $t$. At small $t$ there is an expansion with infinite radius of convergence
\begin{eqnarray}
    \mathcal K(t)=4\sqrt{\frac{t}{\pi}}\left(1+\frac{8 t^2}{15}+\frac{32 t^4}{315}+\frac{256 t^6}{27027}+\frac{512 t^8}{984555}+\frac{4096
   t^{10}}{218243025}+\right) 
   \qquad, \quad t\to 0^+
   \label{eq:wlgn-small-t}
\end{eqnarray}
At large $t$ the expansion along the positive real axis is factorially divergent, with zero radius of convergence:
\begin{eqnarray}
    \mathcal K(t) \sim \frac{e^{2t}}{\sqrt{2\pi t}}\left(1+\frac{3}{16 t}+\frac{57}{512 t^2}+\frac{945}{8192 t^3}+\frac{91035}{524288 t^4}+\frac{2900205}{8388608
   t^5}+\dots \right) \qquad, \quad t\to +\infty
   \label{eq:wlgn-large-t}
\end{eqnarray}
The coefficients appearing in the small $t$ expansion \eqref{eq:wlgn-small-t} are 
\begin{eqnarray}
    c_n&=&
    \frac{\pi}{2^{2n+3/2}}\frac{\Gamma(2n+1)}{(n!)^2 \Gamma(n+3/4)\Gamma(n+5/4)}=
    \left\{1,\frac{8}{15},\frac{32}{315},\frac{256}{27027},\frac{512}{984555},\frac{4096}{218243025}, \dots \right\}
    \label{eq:wlgn-cn}
\end{eqnarray}
The small $t$ expansion coefficients $c_n$ decrease factorially fast, with leading behavior
\begin{eqnarray}
    c_n\sim \frac{2^{2n-1/2}}{(2n+1)!} \qquad, \quad n\to +\infty 
      \label{eq:wlgn-cn-leading}
\end{eqnarray}
The subleading power-law corrections to this leading behavior can be expressed through an expansion in inverse factorials of growing argument:
\begin{eqnarray}
    c_n&\sim& \frac{2^{2n-1/2}}{(2n+1)!}\left[1 +\frac{3\cdot 2}{16\, (2n+2)} +\frac{57\cdot 2^2}{512\, (2n+2)(2n+3)}
    +\frac{945\cdot 2^3}{8192(2n+2)(2n+3)(2n+4)}+ \right.
    \nonumber\\
    &&\hskip 2cm +\left. 
    \frac{91035\cdot 2^4}{524288(2n+2)(2n+3)(2n+4)(2n+5)}+\dots \right] \qquad, \quad n\to +\infty 
      \label{eq:wlgn-cn-subleading}
\end{eqnarray}
The powers of 2 appear due to the leading $2^{2n}$ growth in \eqref{eq:wlgn-cn-leading}.
We recognize the coefficients of the subleading terms of this large order $n\to\infty$ as the same coefficients appearing in the factorially divergent large $t$ expansion \eqref{eq:wlgn-large-t}.
This means that information about the (factorially divergent) $t\to+\infty$ behavior of the heat kernel $K(t)$ is encoded in the (factorially convergent) $t\to0^+$ behavior of $K(t)$. 

\section{Generalized Pearcey Function}
\label{sec:gen-pearcey}

The relations described in sections \ref{sec:p1} and \ref{sec:p2} can be extended in a natural way to more general functions where the $\phi^4$ interaction term  in the exponent becomes $\phi^{2M}$, with $M\in \mathbb{Z}$,  $M\geq 2$.

\subsection{Generalized Pearcey 1: large-order/low-order dualities}
\label{sec:gp1}

Consider the following generalization of the Pearcey 1 function in section \ref{sec:p1}
\begin{eqnarray} 
GP_1(y)&=& \int_0^\infty d\phi\, \exp\left[-\phi^{2M}-y\, \phi\right] \qquad, \qquad M\in \mathbb Z \geq 2\label{eqn:gp1}
\\
&\sim &
\sum_{k=0}^\infty (-1)^k \frac{\Gamma\left(2M k+1\right)}{k!}\frac{1}{y^{2Mk+1}}
\qquad, \quad y\to +\infty
\label{eq:gp1-strong}
\\
&=&
\frac{1}{2M}\sum_{n=0}^\infty (-1)^n \frac{\Gamma\left((n+1)/(2M)\right)}{n!} \, y^n 
\qquad, \quad y\to 0
\label{eq:gp1-weak}
\end{eqnarray} 
The $y\to +\infty$ expansion is factorially divergent [zero radius of convergence], while the $y\to 0$ expansion is factorially convergent [infinite radius of convergence]. Setting $M=2$ reduces to the Pearcey 1 function discussed in section \ref{sec:p1}.

The large order ($k\to \infty$) behavior of the coefficients of the $y\to +\infty$ expansion can be written as a series of decreasing factorials:
\begin{eqnarray}
\frac{\Gamma\left(2M k+1\right)}{\Gamma(k+1)}
&\sim& \sqrt{\frac{M}{\pi}} \Gamma\left((2M-1)k+\frac{1}{2}\right) \left(\frac{(2M)^{2M}}{{(2M-1)^{2M-1}}}\right)^k \left[1- \right.
\nonumber\\
&& \left. \hskip -1cm \frac{(M-1)(4M-1)}{24M\left((2M-1)k-\frac{1}{2}\right)}
+\frac{(M-1)(4M-1)(4M^2+19M+1)}{1152M^2 \left((2M-1)k-\frac{1}{2}\right)\left((2M-1)k-\frac{3}{2}\right)}+O\left(\frac{1}{k^3}\right) \right]
\label{eq:gp1-large-k}
\end{eqnarray}
The large order ($n\to \infty$) behavior of the coefficients of the $y\to 0$ expansion can be written as a series of inverses of increasing factorials:
\begin{eqnarray}
    \frac{1}{2M}\frac{\Gamma\left((n+1)/(2M)\right)}{\Gamma(n+1)} &\sim& \sqrt{\frac{\pi}{M}}
\frac{1}{\Gamma\left((2M-1)\frac{(n+1)}{2M}+\frac{1}{2}\right)}
\left(\frac{(2M-1)^{2M-1}}{(2M)^{2M}}\right)^{(n+1)/(2M)}\left[1+ \right.
\nonumber\\
&& \left. \hskip -3cm \frac{(M-1)(4M-1)}{24M\left((2M-1)\frac{(n+1)}{2M}+\frac{1}{2}\right)}
+\frac{(M-1)(4M-1)(4M^2+19M+1)}{1152M^2 \left((2M-1)\frac{(n+1)}{2M}+\frac{1}{2}\right)\left((2M-1)\frac{(n+1)}{2M}+\frac{3}{2}\right)}+O\left(\frac{1}{n^3}\right) \right]
\label{eq:gp1-large-n}
\end{eqnarray}
From these expansions we observe that the same coefficients appear in the large order behavior in \eqref{eq:gp1-large-k} and \eqref{eq:gp1-large-n}. In other words, information about the factorially divergent expansion at infinity is encoded in the factorially convergent expansion at zero, and vice versa.

In fact, using the gamma function reflection formula 
\begin{eqnarray}
    \Gamma\left(\frac{1}{2}-q\right) \Gamma\left(\frac{1}{2}+q\right)=\frac{\pi}{\cos(\pi q)}
    \label{eq:gamma}
\end{eqnarray}
we see that the large order expansions in \eqref{eq:gp1-large-k} and \eqref{eq:gp1-large-n} map into one another under the mapping:
\begin{eqnarray}
    k\quad \longleftrightarrow \quad -\left(\frac{n+1}{2M}\right)
    \label{eq:gp1-map}
\end{eqnarray}
When $M=2$, this agrees with \eqref{eq:p1-map}. See section \ref{sec:mellin} for another perspective on the mapping \eqref{eq:gp1-map}.

\subsection{Generalized Pearcey 2: 
large-order/low-order dualities}
\label{sec:gp2}

Consider the following generalization of the Pearcey 2 function in section \ref{sec:p2}
\begin{eqnarray}
GP_2(x)&=& \int_0^\infty d\phi\, \exp\left[-\phi^{2M}-x\, \phi^2\right] \qquad, \qquad M\in \mathbb Z \geq 2
\label{eq:gp2}
\\
&\sim &
\frac{1}{2} \sum_{k=0}^\infty (-1)^k \frac{\Gamma\left(M k+\frac{1}{2}\right)}{k!}\frac{1}{x^{M k+1/2}}
\qquad, \quad x\to +\infty
\label{eq:gp2-strong}
\\
&=&
\frac{1}{2M}\sum_{n=0}^\infty (-1)^n \frac{\Gamma\left(\left(n+\frac{1}{2}\right)/M\right)}{n!} \, x^n \qquad, \qquad x\to 0
\label{eq:gp2-weak}
\end{eqnarray}
The $x\to +\infty$ expansion is factorially divergent [zero radius of convergence], while the $x\to 0$ expansion is factorially convergent [infinite radius of convergence]. Setting $M=2$ reduces to the Pearcey 2 function discussed in section \ref{sec:p2}.

The large order behavior of the coefficients of the $x\to +\infty$ expansion can be written as a series of decreasing factorials:
\begin{eqnarray}
\frac{\Gamma\left(M k+\frac{1}{2}\right)}{\Gamma(k+1)}
&\sim& \sqrt{\frac{M-1}{2\pi}} \Gamma\left((M-1)k\right)
\left(\frac{(M)^{M}}{{(M-1)^{M-1}}}\right)^k   \left[1- \frac{(M+1)(2M-1)}{24M} \frac{1}{((M-1)k-1)} 
\right.
\nonumber\\
&& \left.  
+  \frac{(M+1)(2M-1)(2M^2+49M-1)}{1152 M^2 ((M-1)k-1)((M-1)k-2)}+O\left(\frac{1}{k^3}\right)
 \right]
 \label{eq:gp2-large-k}
\end{eqnarray}
The large order behavior of the coefficients of the $x\to 0$ expansion can be written in a series of inverses of increasing factorials:
\begin{eqnarray}
\frac{\Gamma\left(\left(n+\frac{1}{2}\right)/M\right)}{\Gamma(n+1)}
&=&  \frac{\sqrt{2\pi(M-1)}}{\Gamma\left(\left(n+\frac{1}{2}\right)\left(\frac{M-1}{M}\right)+1\right)}\left(\frac{(M-1)^{M-1}}{M^M}\right)^{(n+\frac{1}{2})/M}
\left[1+ \frac{(M+1)(2M-1)}{24M\left(\left(n+\frac{1}{2}\right)\left(\frac{M-1}{M}\right)+1\right)} \right.
\nonumber\\
&&  \left. + \frac{(M+1)(2M-1)(2M^2+49M-1)}{1152M^2\left(\left(n+\frac{1}{2}\right)\left(\frac{M-1}{M}\right)+1\right)\left(\left(n+\frac{1}{2}\right)\left(\frac{M-1}{M}\right)+2\right)}
+O\left(\frac{1}{n^3}\right)\right]
\end{eqnarray}
From these expansions we observe that the same coefficients appear in the large order behavior in \eqref{eq:gp2-large-k}. 

In fact, as before,
we see that these large order expansions map into one another under the mapping:
\begin{eqnarray}
   k\quad \longleftrightarrow \quad -\left(\frac{n+1/2}{M}\right)
    \label{eq:gp2-map}
\end{eqnarray}
When $M=2$, this agrees with \eqref{eq:p2-map}.

\section{Mellin Transform  Asymptotics}
\label{sec:mellin}

In this section we show how the Mellin transform approach to asymptotics \cite{flajolet-book} explains the mappings between the weak and strong coupling expansions for the Pearcey and generalized Pearcey functions in sections \ref{sec:p1}, \ref{sec:p2}, \ref{sec:gp1} and \ref{sec:gp2}. See equations \eqref{eq:p1-map}, \eqref{eq:p2-map}, \eqref{eq:gp1-map} and \eqref{eq:gp2-map}.

\noindent The Mellin transform is an integral transform that is particularly useful in problems with  asymptotic analysis. For a function $f(x)$ defined on $x>0$, its Mellin transform is given by
\begin{equation}
\mathcal{M}\{f\}(s) = \int_0^\infty x^{s-1} f(x)\,dx,
\end{equation}
provided the integral converges in a suitable strip of the complex $s$-plane. The original function can be recovered from its Mellin transform by the inverse Mellin transform,
\begin{equation}
f(x) = \frac{1}{2\pi i}\int_{c-i\infty}^{c+i\infty} x^{-s}\,\mathcal{M}\{f\}(s)\,ds,
\end{equation}
where the contour is a vertical line $\operatorname{Re}(s)=c$ chosen within the domain of analyticity of $\mathcal{M}\{f\}(s)$. In practice, this representation is often evaluated by contour deformation and residue calculus: 
\begin{align}
    \label{eq:inverse-mellin}
    f(x) 
    &= \sum_n \operatorname{Res}_n x^{-s}\mathcal{M}\{f(s)\}.
\end{align}
Poles on the positive real $s$ axis generate the large $x$ expansion, while poles on the negative real $s$ axis generate the small $x$ expansion.

\begin{figure}[h]
\centering
\includegraphics[width=0.7\textwidth]{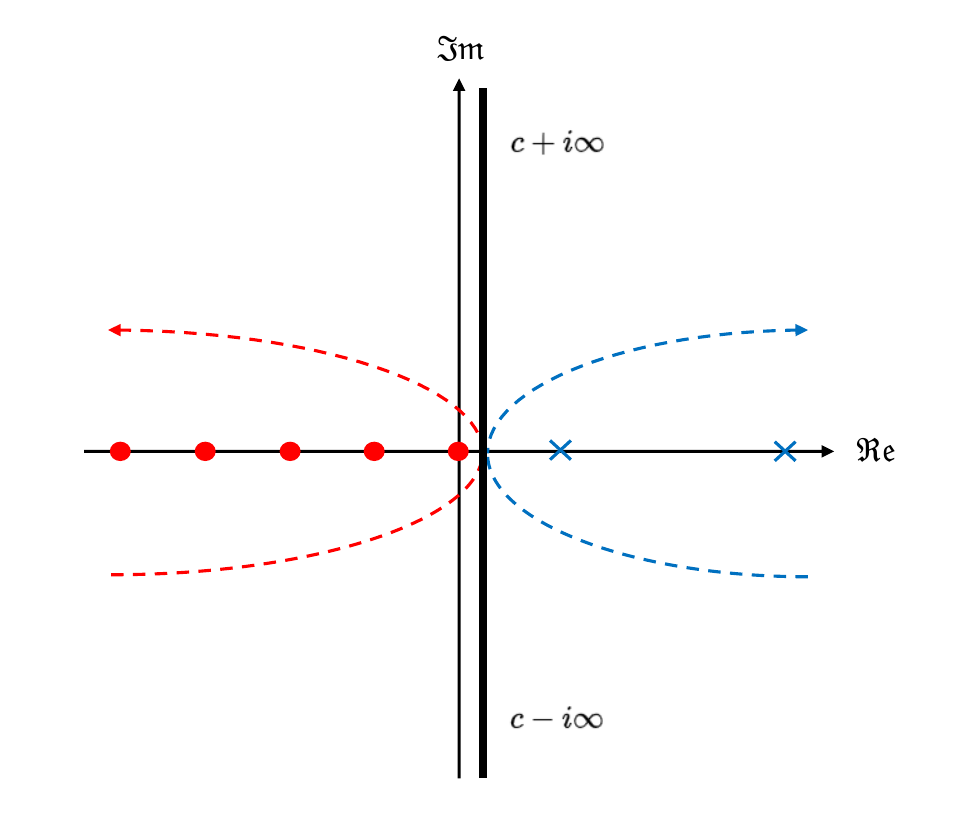}
    \caption{A sketch of the two horseshoe contour deformations for the inverse Mellin transform. The dots and crosses along the real axis denote the poles of \eqref{eq:mellin-p1} coming from the Gamma function. Integrating about the contour $\mathrm{Re}(s) = c$ wrapped to the left (red) recovers the weak-coupling expansion, while wrapping to the right (blue) recovers the strong-coupling expansion for the Pearcey integral.}
    \label{fig:horseshoe-contours}
\end{figure}

For example, consider the Pearcey function $P_1(y)$ discussed in section \ref{sec:p1}. The Mellin transform is a simple product of two gamma functions: 
\begin{equation} 
\mathcal{M}\{P_1\}(s) = \frac{1}{4}\Gamma(s)\Gamma\left(\frac{1-s}{4}\right).
\label{eq:mellin-p1}
\end{equation}
The poles of the $\Gamma\left(\frac{1-s}{4}\right)$ factor lie on the positive real axis at $s=4k+1$, for $k=0, 1, 2, ...$, while the poles of the $\Gamma\left(s\right)$ factor lie   on the negative real axis at $s=-n$, with $n=0, 1, 2, \dots$ (see Figure \ref{fig:horseshoe-contours}).
The contour for the inverse Mellin transform can be taken along the vertical line $(c-i\,\infty, c+i\,\infty)$ with $0<c<1$. Closing the contour to the right produces the large $y$ expansion in \eqref{eq:p1-strong}, while closing the contour to the left produces the small $y$ expansion in \eqref{eq:p1-weak}.
The interchange between these two gamma function factors in the Mellin transform explains the mapping symmetry in \eqref{eq:p1-map}.

Similarly, for the generalized Pearcey function $GP_1$ discussed in section \ref{sec:gp1}. The Mellin transform is a simple product of two gamma functions: 
\begin{equation} 
\mathcal{M}\{GP_1\}(s) = \frac{1}{2M}\Gamma(s)\Gamma\left(\frac{1-s}{2M}\right).
\end{equation}
The poles of the $\Gamma\left(\frac{1-s}{2M}\right)$ factor lie on the positive real axis at $s=2Mk+1$, for $k=0, 1, 2, ...$, while the poles of the $\Gamma\left(s\right)$ factor lie   on the negative real axis at $s=-n$, with $n=0, 1, 2, \dots$. 
The contour for the inverse Mellin transform can be taken along the vertical line $(c-i\,\infty, c+i\,\infty)$ with $0<c<1$. Closing the contour to the right produces the large $y$ expansion in \eqref{eq:gp1-strong}, while closing the contour to the left produces the small $y$ expansion in \eqref{eq:gp1-weak}.
The interchange between these two gamma function factors in the Mellin transform explains the mapping symmetry in \eqref{eq:gp1-map}.

An analogous analysis applies to the second Pearcey function $P_2(x)$ in section \ref{sec:p2}.
For the special case of the second Pearcey integral given by (\ref{eqn:Pearcey2}) whose Mellin transform is:
\begin{equation} \label{Eqn.13}
\mathcal{M}\{P_2\}(s) = \frac{1}{4}\Gamma(s)\Gamma\left(\frac{1}{4}-\frac{s}{2}\right).
\end{equation}
We can recover (\ref{eq:p2-small-x})—the short-time asymptotic expansion of the second Pearcey integral—by taking the inverse Mellin transform of (\ref{Eqn.13}) using the vertical contour $\mathrm{Re} (s) = c$ wrapped around the negative integers, which are the poles of \(\Gamma(s)\). The residues of \(\Gamma(s)\) are given by \[\text{Res}_{s=-n}\Gamma(s)=\frac{(-1)^n}{n!},\] so substituting this into the integrand of (\ref{eq:inverse-mellin}) and letting \(s \to -n\), we recover
\begin{equation}
P_2(x) = \frac{1}{4}\sum_{n=0}^{\infty} (-x)^n \frac{\Gamma\!\left(\tfrac{n}{2}+\tfrac{1}{4}\right)}{\Gamma(n+1)} ,\qquad x>0
\end{equation}
which is precisely (\ref{eq:p2-small-x}). Likewise, we can do the same thing for the contour deformed to wrap around positive \(s\) for the poles of \(\Gamma(1/4-s/2)\), given by
\begin{equation}
s_0 = 2k+\frac{1}{2}; \qquad k=0, 1, 2, ...
\end{equation}
It can be shown that
\begin{equation}
\text{Res}_{s=s_0} \Gamma \left(\frac{1}{4}-\frac{s}{2} \right) =(-1)^{k+1}\frac{2}{k!}.
\end{equation}
Again, by substituting this in for \(\Gamma(1/4-s/2)\) in the integrand of (\ref{eq:inverse-mellin}) and letting \(s \to s_0\), it follows similarly that
\begin{equation}
P_2(x) = \frac{1}{2} \sum^{\infty}_{k=0} (-1)^{k+1} \frac{\Gamma\left(2k+\frac{1}{2}\right)}{k!} \frac{1}{ x^{2k+1/2}}, \qquad x \to \infty.
\end{equation}
This explains the explicit mapping between the small and large expansions in \eqref{eq:p1-map} and (\ref{eq:p2-map}).

\section{Conclusions}
\label{sec:conclusion}

In this paper we have shown that the global continuation from the origin to infinity for a class of functions with factorially decreasing coefficients at the origin, as studied in \cite{costin-xia}, has an additional duality:
the expansions at the origin and at infinity are directly related to one another, and also to the rate of growth of their coefficients.
This goes beyond the well-known large-order/low-order resurgent relations for factorially divergent expansions, which connect the large order growth of the coefficients with the low order coefficients of the asymptotic expansion \cite{berry-howls}.
These dualities have been  illustrated with the Airy and Pearcey functions, the first of the catastrophe integrals, and for generalizations thereof. 
These dualities can also be understood using the Mellin transform approach to asymptotics. Two physical applications have been presented, related to the Green's functions of a zero-dimensional scalar quantum field theory, and to the semiclassical analysis of the Gross-Neveu model.
Further potential applications in QFT could be to study Wilson loops in SUSY QFTs \cite{Drukker_2001}, and mathematically to characterize precisely the most general class of functions for which these new dualities apply.

\section{Acknowledgments}
 This material is based upon work supported by the U.S. Department of Energy, Office of Science, Office of High Energy Physics under Award Number DE-SC0010339 (GD). AF acknowledges Michael Petro and Dr. Ren\'ee Trueman for the CAPS Research Summer internship from the McNair Scholars Program, and the support of the Mark E. Miller Award for summer undergraduate research from the University of Connecticut's Department of Physics.

\bibliography{bibliography}

%apsrev4-2.bst 2019-01-14 (MD) hand-edited version of apsrev4-1.bst
%Control: key (0)
%Control: author (8) initials jnrlst
%Control: editor formatted (1) identically to author
%Control: production of article title (0) allowed
%Control: page (0) single
%Control: year (1) truncated
%Control: production of eprint (0) enabled
\begin{thebibliography}{35}%
\makeatletter
\providecommand \@ifxundefined [1]{%
 \@ifx{#1\undefined}
}%
\providecommand \@ifnum [1]{%
 \ifnum #1\expandafter \@firstoftwo
 \else \expandafter \@secondoftwo
 \fi
}%
\providecommand \@ifx [1]{%
 \ifx #1\expandafter \@firstoftwo
 \else \expandafter \@secondoftwo
 \fi
}%
\providecommand \natexlab [1]{#1}%
\providecommand \enquote  [1]{``#1''}%
\providecommand \bibnamefont  [1]{#1}%
\providecommand \bibfnamefont [1]{#1}%
\providecommand \citenamefont [1]{#1}%
\providecommand \href@noop [0]{\@secondoftwo}%
\providecommand \href [0]{\begingroup \@sanitize@url \@href}%
\providecommand \@href[1]{\@@startlink{#1}\@@href}%
\providecommand \@@href[1]{\endgroup#1\@@endlink}%
\providecommand \@sanitize@url [0]{\catcode `\\12\catcode `\$12\catcode
  `\&12\catcode `\#12\catcode `\^12\catcode `\_12\catcode `\%12\relax}%
\providecommand \@@startlink[1]{}%
\providecommand \@@endlink[0]{}%
\providecommand \url  [0]{\begingroup\@sanitize@url \@url }%
\providecommand \@url [1]{\endgroup\@href {#1}{\urlprefix }}%
\providecommand \urlprefix  [0]{URL }%
\providecommand \Eprint [0]{\href }%
\providecommand \doibase [0]{https://doi.org/}%
\providecommand \selectlanguage [0]{\@gobble}%
\providecommand \bibinfo  [0]{\@secondoftwo}%
\providecommand \bibfield  [0]{\@secondoftwo}%
\providecommand \translation [1]{[#1]}%
\providecommand \BibitemOpen [0]{}%
\providecommand \bibitemStop [0]{}%
\providecommand \bibitemNoStop [0]{.\EOS\space}%
\providecommand \EOS [0]{\spacefactor3000\relax}%
\providecommand \BibitemShut  [1]{\csname bibitem#1\endcsname}%
\let\auto@bib@innerbib\@empty
%</preamble>
\bibitem [{\citenamefont {\'Ecalle}(1993)}]{ecalle3}%
  \BibitemOpen
  \bibfield  {author} {\bibinfo {author} {\bibfnamefont {J.}~\bibnamefont
  {\'Ecalle}},\ }\bibinfo {title} {In: {\it Bifurcations and Periodic Orbits of
  Vector Fields}}\ (\bibinfo  {publisher} {Springer},\ \bibinfo {year} {1993})\
  Chap.\ \bibinfo {chapter} {{Six Lectures on Transseries, Analysable Functions
  and the Constructive Proof of Dulac’s Conjecture}}\BibitemShut {NoStop}%
\bibitem [{\citenamefont {Mitschi}\ and\ \citenamefont
  {Sauzin}(2016)}]{sauzin}%
  \BibitemOpen
  \bibfield  {author} {\bibinfo {author} {\bibfnamefont {C.}~\bibnamefont
  {Mitschi}}\ and\ \bibinfo {author} {\bibfnamefont {D.}~\bibnamefont
  {Sauzin}},\ }\href@noop {} {\emph {\bibinfo {title} {Divergent series,
  summability and resurgence}}}\ (\bibinfo  {publisher} {Springer},\ \bibinfo
  {year} {2016})\BibitemShut {NoStop}%
\bibitem [{\citenamefont {C\'{\i}\ifmmode~\check{z}\else \v{z}\fi{}ek}\ \emph
  {et~al.}(1986)\citenamefont {C\'{\i}\ifmmode~\check{z}\else \v{z}\fi{}ek},
  \citenamefont {Damburg}, \citenamefont {Graffi}, \citenamefont {Grecchi},
  \citenamefont {Harrell}, \citenamefont {Harris}, \citenamefont {Nakai},
  \citenamefont {Paldus}, \citenamefont {Propin},\ and\ \citenamefont
  {Silverstone}}]{h2plus}%
  \BibitemOpen
  \bibfield  {author} {\bibinfo {author} {\bibfnamefont {J.~c.~v.}\
  \bibnamefont {C\'{\i}\ifmmode~\check{z}\else \v{z}\fi{}ek}}, \bibinfo
  {author} {\bibfnamefont {R.~J.}\ \bibnamefont {Damburg}}, \bibinfo {author}
  {\bibfnamefont {S.}~\bibnamefont {Graffi}}, \bibinfo {author} {\bibfnamefont
  {V.}~\bibnamefont {Grecchi}}, \bibinfo {author} {\bibfnamefont {E.~M.}\
  \bibnamefont {Harrell}}, \bibinfo {author} {\bibfnamefont {J.~G.}\
  \bibnamefont {Harris}}, \bibinfo {author} {\bibfnamefont {S.}~\bibnamefont
  {Nakai}}, \bibinfo {author} {\bibfnamefont {J.}~\bibnamefont {Paldus}},
  \bibinfo {author} {\bibfnamefont {R.~K.}\ \bibnamefont {Propin}},\ and\
  \bibinfo {author} {\bibfnamefont {H.~J.}\ \bibnamefont {Silverstone}},\
  }\bibfield  {title} {\bibinfo {title} {{1/R expansion for
  ${\mathrm{H}}_{2}$${\mathrm{}}^{+}$: Calculation of exponentially small terms
  and asymptotics}},\ }\href {https://doi.org/10.1103/PhysRevA.33.12}
  {\bibfield  {journal} {\bibinfo  {journal} {Phys. Rev. A}\ }\textbf {\bibinfo
  {volume} {33}},\ \bibinfo {pages} {12} (\bibinfo {year} {1986})}\BibitemShut
  {NoStop}%
\bibitem [{\citenamefont {Alvarez}\ \emph {et~al.}(2002)\citenamefont
  {Alvarez}, \citenamefont {Howls},\ and\ \citenamefont
  {Silverstone}}]{alvarez}%
  \BibitemOpen
  \bibfield  {author} {\bibinfo {author} {\bibfnamefont {G.}~\bibnamefont
  {Alvarez}}, \bibinfo {author} {\bibfnamefont {C.~J.}\ \bibnamefont {Howls}},\
  and\ \bibinfo {author} {\bibfnamefont {H.~J.}\ \bibnamefont {Silverstone}},\
  }\bibfield  {title} {\bibinfo {title} {{Anharmonic oscillator discontinuity
  formulae up to second-exponentially-small order}},\ }\href@noop {} {\bibfield
   {journal} {\bibinfo  {journal} {J. Phys. A: Math. Gen.}\ }\textbf {\bibinfo
  {volume} {35}},\ \bibinfo {pages} {4003} (\bibinfo {year}
  {2002})}\BibitemShut {NoStop}%
\bibitem [{\citenamefont {Zinn-Justin}\ and\ \citenamefont
  {Jentschura}(2004{\natexlab{a}})}]{Zinn-Justin:2004vcw}%
  \BibitemOpen
  \bibfield  {author} {\bibinfo {author} {\bibfnamefont {J.}~\bibnamefont
  {Zinn-Justin}}\ and\ \bibinfo {author} {\bibfnamefont {U.~D.}\ \bibnamefont
  {Jentschura}},\ }\bibfield  {title} {\bibinfo {title} {{Multi-instantons and
  exact results I: Conjectures, WKB expansions, and instanton interactions}},\
  }\href {https://doi.org/10.1016/j.aop.2004.04.004} {\bibfield  {journal}
  {\bibinfo  {journal} {Annals Phys.}\ }\textbf {\bibinfo {volume} {313}},\
  \bibinfo {pages} {197} (\bibinfo {year} {2004}{\natexlab{a}})},\ \Eprint
  {https://arxiv.org/abs/quant-ph/0501136} {arXiv:quant-ph/0501136}
  \BibitemShut {NoStop}%
\bibitem [{\citenamefont {Zinn-Justin}\ and\ \citenamefont
  {Jentschura}(2004{\natexlab{b}})}]{Zinn-Justin:2004qzw}%
  \BibitemOpen
  \bibfield  {author} {\bibinfo {author} {\bibfnamefont {J.}~\bibnamefont
  {Zinn-Justin}}\ and\ \bibinfo {author} {\bibfnamefont {U.~D.}\ \bibnamefont
  {Jentschura}},\ }\bibfield  {title} {\bibinfo {title} {{Multi-instantons and
  exact results II: Specific cases, higher-order effects, and numerical
  calculations}},\ }\href {https://doi.org/10.1016/j.aop.2004.04.003}
  {\bibfield  {journal} {\bibinfo  {journal} {Annals Phys.}\ }\textbf {\bibinfo
  {volume} {313}},\ \bibinfo {pages} {269} (\bibinfo {year}
  {2004}{\natexlab{b}})},\ \Eprint {https://arxiv.org/abs/quant-ph/0501137}
  {arXiv:quant-ph/0501137} \BibitemShut {NoStop}%
\bibitem [{\citenamefont {Mari{\~n}o}\ \emph {et~al.}(2008)\citenamefont
  {Mari{\~n}o}, \citenamefont {Schiappa},\ and\ \citenamefont
  {Weiss}}]{Marino:2007te}%
  \BibitemOpen
  \bibfield  {author} {\bibinfo {author} {\bibfnamefont {M.}~\bibnamefont
  {Mari{\~n}o}}, \bibinfo {author} {\bibfnamefont {R.}~\bibnamefont
  {Schiappa}},\ and\ \bibinfo {author} {\bibfnamefont {M.}~\bibnamefont
  {Weiss}},\ }\bibfield  {title} {\bibinfo {title} {{Nonperturbative Effects
  and the Large-Order Behavior of Matrix Models and Topological Strings}},\
  }\href {https://doi.org/10.4310/CNTP.2008.v2.n2.a3} {\bibfield  {journal}
  {\bibinfo  {journal} {Commun. Num. Theor. Phys.}\ }\textbf {\bibinfo {volume}
  {2}},\ \bibinfo {pages} {349} (\bibinfo {year} {2008})},\ \Eprint
  {https://arxiv.org/abs/0711.1954} {arXiv:0711.1954 [hep-th]} \BibitemShut
  {NoStop}%
\bibitem [{\citenamefont {Mari{\~n}o}(2014)}]{Marino:2012zq}%
  \BibitemOpen
  \bibfield  {author} {\bibinfo {author} {\bibfnamefont {M.}~\bibnamefont
  {Mari{\~n}o}},\ }\bibfield  {title} {\bibinfo {title} {{Lectures on
  non-perturbative effects in large $N$ gauge theories, matrix models and
  strings}},\ }\href {https://doi.org/10.1002/prop.201400005} {\bibfield
  {journal} {\bibinfo  {journal} {Fortsch. Phys.}\ }\textbf {\bibinfo {volume}
  {62}},\ \bibinfo {pages} {455} (\bibinfo {year} {2014})},\ \Eprint
  {https://arxiv.org/abs/1206.6272} {arXiv:1206.6272 [hep-th]} \BibitemShut
  {NoStop}%
\bibitem [{\citenamefont {Dunne}\ and\ \citenamefont
  {{\"U}nsal}(2016)}]{annrev}%
  \BibitemOpen
  \bibfield  {author} {\bibinfo {author} {\bibfnamefont {G.~V.}\ \bibnamefont
  {Dunne}}\ and\ \bibinfo {author} {\bibfnamefont {M.}~\bibnamefont
  {{\"U}nsal}},\ }\bibfield  {title} {\bibinfo {title} {{New Nonperturbative
  Methods in Quantum Field Theory: From Large-N Orbifold Equivalence to Bions
  and Resurgence}},\ }\href
  {https://doi.org/10.1146/annurev-nucl-102115-044755} {\bibfield  {journal}
  {\bibinfo  {journal} {Ann. Rev. Nucl. Part. Sci.}\ }\textbf {\bibinfo
  {volume} {66}},\ \bibinfo {pages} {245} (\bibinfo {year} {2016})},\ \Eprint
  {https://arxiv.org/abs/1601.03414} {arXiv:1601.03414 [hep-th]} \BibitemShut
  {NoStop}%
\bibitem [{\citenamefont {Aniceto}\ \emph {et~al.}(2019)\citenamefont
  {Aniceto}, \citenamefont {Ba{\c s}ar},\ and\ \citenamefont
  {Schiappa}}]{aniceto}%
  \BibitemOpen
  \bibfield  {author} {\bibinfo {author} {\bibfnamefont {I.}~\bibnamefont
  {Aniceto}}, \bibinfo {author} {\bibfnamefont {G.}~\bibnamefont {Ba{\c
  s}ar}},\ and\ \bibinfo {author} {\bibfnamefont {R.}~\bibnamefont
  {Schiappa}},\ }\bibfield  {title} {\bibinfo {title} {{A Primer on Resurgent
  Transseries and Their Asymptotics}},\ }\href
  {https://doi.org/10.1016/j.physrep.2019.02.003} {\bibfield  {journal}
  {\bibinfo  {journal} {Phys. Rept.}\ }\textbf {\bibinfo {volume} {809}},\
  \bibinfo {pages} {1} (\bibinfo {year} {2019})},\ \Eprint
  {https://arxiv.org/abs/1802.10441} {arXiv:1802.10441 [hep-th]} \BibitemShut
  {NoStop}%
\bibitem [{\citenamefont {Berry}\ and\ \citenamefont
  {Howls}(1991)}]{berry-howls}%
  \BibitemOpen
  \bibfield  {author} {\bibinfo {author} {\bibfnamefont {M.~V.}\ \bibnamefont
  {Berry}}\ and\ \bibinfo {author} {\bibfnamefont {C.~J.}\ \bibnamefont
  {Howls}},\ }\bibfield  {title} {\bibinfo {title} {{Hyperasymptotics for
  integrals with saddles}},\ }\href {https://doi.org/10.1098/rspa.1991.0119}
  {\bibfield  {journal} {\bibinfo  {journal} {Proc. Roy. Soc. Lond. A}\
  }\textbf {\bibinfo {volume} {434}},\ \bibinfo {pages} {657} (\bibinfo {year}
  {1991})}\BibitemShut {NoStop}%
\bibitem [{\citenamefont {Costin}\ and\ \citenamefont {Dunne}(2022)}]{cd}%
  \BibitemOpen
  \bibfield  {author} {\bibinfo {author} {\bibfnamefont {O.}~\bibnamefont
  {Costin}}\ and\ \bibinfo {author} {\bibfnamefont {G.~V.}\ \bibnamefont
  {Dunne}},\ }\bibfield  {title} {\bibinfo {title} {{Uniformization and
  Constructive Analytic Continuation of Taylor Series}},\ }\href
  {https://doi.org/10.1007/s00220-022-04361-6} {\bibfield  {journal} {\bibinfo
  {journal} {Commun. Math. Phys.}\ }\textbf {\bibinfo {volume} {392}},\
  \bibinfo {pages} {863} (\bibinfo {year} {2022})},\ \Eprint
  {https://arxiv.org/abs/2009.01962} {arXiv:2009.01962 [math.CV]} \BibitemShut
  {NoStop}%
\bibitem [{\citenamefont {Guillou}\ and\ \citenamefont
  {Zinn-Justin}(1990)}]{leguillou}%
  \BibitemOpen
  \bibfield  {author} {\bibinfo {author} {\bibfnamefont {J.~L.}\ \bibnamefont
  {Guillou}}\ and\ \bibinfo {author} {\bibfnamefont {J.}~\bibnamefont
  {Zinn-Justin}},\ }\href@noop {} {\emph {\bibinfo {title} {{Large-Order
  Behaviour of Perturbation Theory}}}}\ (\bibinfo  {publisher}
  {North-Holland},\ \bibinfo {year} {1990})\BibitemShut {NoStop}%
\bibitem [{\citenamefont {Gukov}\ \emph {et~al.}(2016)\citenamefont {Gukov},
  \citenamefont {Mari{\~n}o},\ and\ \citenamefont {Putrov}}]{Gukov:2016njj}%
  \BibitemOpen
  \bibfield  {author} {\bibinfo {author} {\bibfnamefont {S.}~\bibnamefont
  {Gukov}}, \bibinfo {author} {\bibfnamefont {M.}~\bibnamefont {Mari{\~n}o}},\
  and\ \bibinfo {author} {\bibfnamefont {P.}~\bibnamefont {Putrov}},\
  }\bibfield  {title} {\bibinfo {title} {{Resurgence in complex Chern-Simons
  theory}},\ }\href@noop {} {\  (\bibinfo {year} {2016})},\ \Eprint
  {https://arxiv.org/abs/1605.07615} {arXiv:1605.07615 [hep-th]} \BibitemShut
  {NoStop}%
\bibitem [{\citenamefont {Costin}\ \emph {et~al.}(2023)\citenamefont {Costin},
  \citenamefont {Dunne}, \citenamefont {Gruen},\ and\ \citenamefont
  {Gukov}}]{Costin:2023kla}%
  \BibitemOpen
  \bibfield  {author} {\bibinfo {author} {\bibfnamefont {O.}~\bibnamefont
  {Costin}}, \bibinfo {author} {\bibfnamefont {G.~V.}\ \bibnamefont {Dunne}},
  \bibinfo {author} {\bibfnamefont {A.}~\bibnamefont {Gruen}},\ and\ \bibinfo
  {author} {\bibfnamefont {S.}~\bibnamefont {Gukov}},\ }\bibfield  {title}
  {\bibinfo {title} {{Going to the Other Side via the Resurgent Bridge}},\
  }\href@noop {} {\  (\bibinfo {year} {2023})},\ \Eprint
  {https://arxiv.org/abs/2310.12317} {arXiv:2310.12317 [hep-th]} \BibitemShut
  {NoStop}%
\bibitem [{\citenamefont {Adams}\ \emph {et~al.}(2025)\citenamefont {Adams},
  \citenamefont {Costin}, \citenamefont {Dunne}, \citenamefont {Gukov},\ and\
  \citenamefont {{\"O}ner}}]{Adams:2025aad}%
  \BibitemOpen
  \bibfield  {author} {\bibinfo {author} {\bibfnamefont {G.}~\bibnamefont
  {Adams}}, \bibinfo {author} {\bibfnamefont {O.}~\bibnamefont {Costin}},
  \bibinfo {author} {\bibfnamefont {G.~V.}\ \bibnamefont {Dunne}}, \bibinfo
  {author} {\bibfnamefont {S.}~\bibnamefont {Gukov}},\ and\ \bibinfo {author}
  {\bibfnamefont {O.}~\bibnamefont {{\"O}ner}},\ }\bibfield  {title} {\bibinfo
  {title} {{Orientation reversal and the Chern-Simons natural boundary}},\
  }\href {https://doi.org/10.1007/JHEP08(2025)154} {\bibfield  {journal}
  {\bibinfo  {journal} {JHEP}\ }\textbf {\bibinfo {volume} {08}},\ \bibinfo
  {pages} {154}},\ \Eprint {https://arxiv.org/abs/2505.14441} {arXiv:2505.14441
  [hep-th]} \BibitemShut {NoStop}%
\bibitem [{\citenamefont {Cheng}\ \emph {et~al.}(2019)\citenamefont {Cheng},
  \citenamefont {Chun}, \citenamefont {Ferrari}, \citenamefont {Gukov},\ and\
  \citenamefont {Harrison}}]{Cheng:2018vpl}%
  \BibitemOpen
  \bibfield  {author} {\bibinfo {author} {\bibfnamefont {M.~C.~N.}\
  \bibnamefont {Cheng}}, \bibinfo {author} {\bibfnamefont {S.}~\bibnamefont
  {Chun}}, \bibinfo {author} {\bibfnamefont {F.}~\bibnamefont {Ferrari}},
  \bibinfo {author} {\bibfnamefont {S.}~\bibnamefont {Gukov}},\ and\ \bibinfo
  {author} {\bibfnamefont {S.~M.}\ \bibnamefont {Harrison}},\ }\bibfield
  {title} {\bibinfo {title} {{3d Modularity}},\ }\href
  {https://doi.org/10.1007/JHEP10(2019)010} {\bibfield  {journal} {\bibinfo
  {journal} {JHEP}\ }\textbf {\bibinfo {volume} {10}},\ \bibinfo {pages}
  {010}},\ \Eprint {https://arxiv.org/abs/1809.10148} {arXiv:1809.10148
  [hep-th]} \BibitemShut {NoStop}%
\bibitem [{\citenamefont {Cheng}\ \emph {et~al.}(2024)\citenamefont {Cheng},
  \citenamefont {Coman}, \citenamefont {Kucharski}, \citenamefont {Passaro},\
  and\ \citenamefont {Sgroi}}]{Cheng:2024vou}%
  \BibitemOpen
  \bibfield  {author} {\bibinfo {author} {\bibfnamefont {M.~C.~N.}\
  \bibnamefont {Cheng}}, \bibinfo {author} {\bibfnamefont {I.}~\bibnamefont
  {Coman}}, \bibinfo {author} {\bibfnamefont {P.}~\bibnamefont {Kucharski}},
  \bibinfo {author} {\bibfnamefont {D.}~\bibnamefont {Passaro}},\ and\ \bibinfo
  {author} {\bibfnamefont {G.}~\bibnamefont {Sgroi}},\ }\bibfield  {title}
  {\bibinfo {title} {{3d Modularity Revisited}},\ }\href@noop {} {\  (\bibinfo
  {year} {2024})},\ \Eprint {https://arxiv.org/abs/2403.14920}
  {arXiv:2403.14920 [hep-th]} \BibitemShut {NoStop}%
\bibitem [{\citenamefont {Fantini}\ and\ \citenamefont
  {Rella}(2024)}]{Fantini:2024ihf}%
  \BibitemOpen
  \bibfield  {author} {\bibinfo {author} {\bibfnamefont {V.}~\bibnamefont
  {Fantini}}\ and\ \bibinfo {author} {\bibfnamefont {C.}~\bibnamefont
  {Rella}},\ }\bibfield  {title} {\bibinfo {title} {{Modular resurgent
  structures}},\ }\href@noop {} {\  (\bibinfo {year} {2024})},\ \Eprint
  {https://arxiv.org/abs/2404.11550} {arXiv:2404.11550 [math.NT]} \BibitemShut
  {NoStop}%
\bibitem [{\citenamefont {Bender}\ \emph {et~al.}(1989)\citenamefont {Bender},
  \citenamefont {Cooper},\ and\ \citenamefont {Simmons}}]{bender-nonunique}%
  \BibitemOpen
  \bibfield  {author} {\bibinfo {author} {\bibfnamefont {C.~M.}\ \bibnamefont
  {Bender}}, \bibinfo {author} {\bibfnamefont {F.}~\bibnamefont {Cooper}},\
  and\ \bibinfo {author} {\bibfnamefont {L.~M.}\ \bibnamefont {Simmons}},\
  }\bibfield  {title} {\bibinfo {title} {{Nonunique Solution to the
  Schwinger-dyson Equations}},\ }\href
  {https://doi.org/10.1103/PhysRevD.39.2343} {\bibfield  {journal} {\bibinfo
  {journal} {Phys. Rev. D}\ }\textbf {\bibinfo {volume} {39}},\ \bibinfo
  {pages} {2343} (\bibinfo {year} {1989})}\BibitemShut {NoStop}%
\bibitem [{\citenamefont {Bender}\ \emph {et~al.}(2023)\citenamefont {Bender},
  \citenamefont {Karapoulitidis},\ and\ \citenamefont
  {Klevansky}}]{Bender:2023ttu}%
  \BibitemOpen
  \bibfield  {author} {\bibinfo {author} {\bibfnamefont {C.~M.}\ \bibnamefont
  {Bender}}, \bibinfo {author} {\bibfnamefont {C.}~\bibnamefont
  {Karapoulitidis}},\ and\ \bibinfo {author} {\bibfnamefont {S.~P.}\
  \bibnamefont {Klevansky}},\ }\bibfield  {title} {\bibinfo {title}
  {{Dyson-Schwinger equations in zero dimensions and polynomial
  approximations}},\ }\href {https://doi.org/10.1103/PhysRevD.108.056002}
  {\bibfield  {journal} {\bibinfo  {journal} {Phys. Rev. D}\ }\textbf {\bibinfo
  {volume} {108}},\ \bibinfo {pages} {056002} (\bibinfo {year} {2023})},\
  \Eprint {https://arxiv.org/abs/2307.01008} {arXiv:2307.01008 [math-ph]}
  \BibitemShut {NoStop}%
\bibitem [{\citenamefont {Gross}\ and\ \citenamefont
  {Neveu}(1974)}]{Gross:1974jv}%
  \BibitemOpen
  \bibfield  {author} {\bibinfo {author} {\bibfnamefont {D.~J.}\ \bibnamefont
  {Gross}}\ and\ \bibinfo {author} {\bibfnamefont {A.}~\bibnamefont {Neveu}},\
  }\bibfield  {title} {\bibinfo {title} {{Dynamical Symmetry Breaking in
  Asymptotically Free Field Theories}},\ }\href
  {https://doi.org/10.1103/PhysRevD.10.3235} {\bibfield  {journal} {\bibinfo
  {journal} {Phys. Rev. D}\ }\textbf {\bibinfo {volume} {10}},\ \bibinfo
  {pages} {3235} (\bibinfo {year} {1974})}\BibitemShut {NoStop}%
\bibitem [{\citenamefont {Thies}(2006)}]{Thies:2006ti}%
  \BibitemOpen
  \bibfield  {author} {\bibinfo {author} {\bibfnamefont {M.}~\bibnamefont
  {Thies}},\ }\bibfield  {title} {\bibinfo {title} {{From relativistic quantum
  fields to condensed matter and back again: Updating the Gross-Neveu phase
  diagram}},\ }\href {https://doi.org/10.1088/0305-4470/39/41/S04} {\bibfield
  {journal} {\bibinfo  {journal} {J. Phys. A}\ }\textbf {\bibinfo {volume}
  {39}},\ \bibinfo {pages} {12707} (\bibinfo {year} {2006})},\ \Eprint
  {https://arxiv.org/abs/hep-th/0601049} {arXiv:hep-th/0601049} \BibitemShut
  {NoStop}%
\bibitem [{\citenamefont {Ba{\c s}ar}\ \emph {et~al.}(2009)\citenamefont {Ba{\c
  s}ar}, \citenamefont {Dunne},\ and\ \citenamefont {Thies}}]{bdt}%
  \BibitemOpen
  \bibfield  {author} {\bibinfo {author} {\bibfnamefont {G.}~\bibnamefont
  {Ba{\c s}ar}}, \bibinfo {author} {\bibfnamefont {G.~V.}\ \bibnamefont
  {Dunne}},\ and\ \bibinfo {author} {\bibfnamefont {M.}~\bibnamefont {Thies}},\
  }\bibfield  {title} {\bibinfo {title} {{Inhomogeneous Condensates in the
  Thermodynamics of the Chiral NJL(2) model}},\ }\href
  {https://doi.org/10.1103/PhysRevD.79.105012} {\bibfield  {journal} {\bibinfo
  {journal} {Phys. Rev. D}\ }\textbf {\bibinfo {volume} {79}},\ \bibinfo
  {pages} {105012} (\bibinfo {year} {2009})},\ \Eprint
  {https://arxiv.org/abs/0903.1868} {arXiv:0903.1868 [hep-th]} \BibitemShut
  {NoStop}%
\bibitem [{\citenamefont {Dunne}\ \emph {et~al.}(2009)\citenamefont {Dunne},
  \citenamefont {Gies}, \citenamefont {Klingmuller},\ and\ \citenamefont
  {Langfeld}}]{Dunne:2009zz}%
  \BibitemOpen
  \bibfield  {author} {\bibinfo {author} {\bibfnamefont {G.}~\bibnamefont
  {Dunne}}, \bibinfo {author} {\bibfnamefont {H.}~\bibnamefont {Gies}},
  \bibinfo {author} {\bibfnamefont {K.}~\bibnamefont {Klingmuller}},\ and\
  \bibinfo {author} {\bibfnamefont {K.}~\bibnamefont {Langfeld}},\ }\bibfield
  {title} {\bibinfo {title} {{Worldline Monte Carlo for fermion models at large
  N(f)}},\ }\href {https://doi.org/10.1088/1126-6708/2009/08/010} {\bibfield
  {journal} {\bibinfo  {journal} {JHEP}\ }\textbf {\bibinfo {volume} {08}},\
  \bibinfo {pages} {010}},\ \Eprint {https://arxiv.org/abs/0903.4421}
  {arXiv:0903.4421 [hep-th]} \BibitemShut {NoStop}%
\bibitem [{\citenamefont {Costin}\ and\ \citenamefont
  {Xia}(2015)}]{costin-xia}%
  \BibitemOpen
  \bibfield  {author} {\bibinfo {author} {\bibfnamefont {O.}~\bibnamefont
  {Costin}}\ and\ \bibinfo {author} {\bibfnamefont {X.}~\bibnamefont {Xia}},\
  }\bibfield  {title} {\bibinfo {title} {{From the Taylor series of analytic
  functions to their global analysis}},\ }\href
  {https://doi.org/10.1016/j.na.2014.08.014} {\bibfield  {journal} {\bibinfo
  {journal} {Nonlinear Analysis}\ }\textbf {\bibinfo {volume} {119}},\ \bibinfo
  {pages} {106} (\bibinfo {year} {2015})}\BibitemShut {NoStop}%
\bibitem [{\citenamefont {Airy}(1838)}]{Airy:1838:ILN}%
  \BibitemOpen
  \bibfield  {author} {\bibinfo {author} {\bibfnamefont {G.~B.}\ \bibnamefont
  {Airy}},\ }\bibfield  {title} {\bibinfo {title} {On the intensity of light in
  the neighbourhood of a caustic},\ }\href@noop {} {\bibfield  {journal}
  {\bibinfo  {journal} {Trans. Camb. Phil. Soc.}\ }\textbf {\bibinfo {volume}
  {6}},\ \bibinfo {pages} {379–402} (\bibinfo {year} {1838})}\BibitemShut
  {NoStop}%
\bibitem [{\citenamefont {Stokes}(1856)}]{stokes1}%
  \BibitemOpen
  \bibfield  {author} {\bibinfo {author} {\bibfnamefont {G.~G.}\ \bibnamefont
  {Stokes}},\ }\bibfield  {title} {\bibinfo {title} {On the numerical
  calculation of a class of definite integrals and infinite series},\
  }\href@noop {} {\bibfield  {journal} {\bibinfo  {journal} {Trans. Camb. Phil.
  Soc.}\ }\textbf {\bibinfo {volume} {9}},\ \bibinfo {pages} {166} (\bibinfo
  {year} {1856})}\BibitemShut {NoStop}%
\bibitem [{\citenamefont {Stokes}(1864)}]{stokes2}%
  \BibitemOpen
  \bibfield  {author} {\bibinfo {author} {\bibfnamefont {G.~G.}\ \bibnamefont
  {Stokes}},\ }\bibfield  {title} {\bibinfo {title} {On the discontinuity of
  arbitrary constants which appear in divergent developments},\ }\href@noop {}
  {\bibfield  {journal} {\bibinfo  {journal} {Trans. Camb. Phil. Soc.}\
  }\textbf {\bibinfo {volume} {10}},\ \bibinfo {pages} {105} (\bibinfo {year}
  {1864})}\BibitemShut {NoStop}%
\bibitem [{\citenamefont {Berry}\ and\ \citenamefont
  {Howls}(1993)}]{berry-airy}%
  \BibitemOpen
  \bibfield  {author} {\bibinfo {author} {\bibfnamefont {M.}~\bibnamefont
  {Berry}}\ and\ \bibinfo {author} {\bibfnamefont {C.}~\bibnamefont {Howls}},\
  }\bibfield  {title} {\bibinfo {title} {{Infinity Interpreted}},\ }\href@noop
  {} {\bibfield  {journal} {\bibinfo  {journal} {Physics World}\ }\textbf
  {\bibinfo {volume} {June}},\ \bibinfo {pages} {35} (\bibinfo {year}
  {1993})}\BibitemShut {NoStop}%
\bibitem [{\citenamefont {Dunne}(2025)}]{dunne-cern}%
  \BibitemOpen
  \bibfield  {author} {\bibinfo {author} {\bibfnamefont {G.~V.}\ \bibnamefont
  {Dunne}},\ }\bibfield  {title} {\bibinfo {title} {{Introductory Lectures on
  Resurgence: CERN Summer School 2024}},\ }\bibfield  {journal} {\bibinfo
  {journal} {The European Physical Journal Special Topics}\ }\href
  {https://doi.org/10.1140/epjs/s11734-026-02135-y}
  {10.1140/epjs/s11734-026-02135-y} (\bibinfo {year} {2025}),\ \Eprint
  {https://arxiv.org/abs/2511.15528} {arXiv:2511.15528 [hep-th]} \BibitemShut
  {NoStop}%
\bibitem [{\citenamefont {Henrici}(1977)}]{henrici}%
  \BibitemOpen
  \bibfield  {author} {\bibinfo {author} {\bibfnamefont {P.}~\bibnamefont
  {Henrici}},\ }\href@noop {} {\emph {\bibinfo {title} {{Applied and
  computational complex analysis. Volume 2}}}}\ (\bibinfo  {publisher} {John
  Wiley \& Sons},\ \bibinfo {year} {1977})\BibitemShut {NoStop}%
\bibitem [{\citenamefont {Paris}(1991)}]{paris}%
  \BibitemOpen
  \bibfield  {author} {\bibinfo {author} {\bibfnamefont {R.~B.}\ \bibnamefont
  {Paris}},\ }\bibfield  {title} {\bibinfo {title} {{The asymptotic behaviour
  of Pearcey's integral for complex variables}},\ }\href
  {https://doi.org/10.1098/rspa.1991.0023} {\bibfield  {journal} {\bibinfo
  {journal} {Proc. R. Soc. Lond.}\ }\textbf {\bibinfo {volume} {432}},\
  \bibinfo {pages} {391} (\bibinfo {year} {1991})}\BibitemShut {NoStop}%
\bibitem [{\citenamefont {Flajolet}\ and\ \citenamefont
  {Sedgewick}(2009)}]{flajolet-book}%
  \BibitemOpen
  \bibfield  {author} {\bibinfo {author} {\bibfnamefont {P.}~\bibnamefont
  {Flajolet}}\ and\ \bibinfo {author} {\bibfnamefont {R.}~\bibnamefont
  {Sedgewick}},\ }\href {https://doi.org/10.1017/CBO9780511801655} {\emph
  {\bibinfo {title} {{Analytic Combinatorics}}}}\ (\bibinfo  {publisher}
  {Cambridge University Press},\ \bibinfo {year} {2009})\BibitemShut {NoStop}%
\bibitem [{\citenamefont {Drukker}\ and\ \citenamefont
  {Gross}(2001)}]{Drukker_2001}%
  \BibitemOpen
  \bibfield  {author} {\bibinfo {author} {\bibfnamefont {N.}~\bibnamefont
  {Drukker}}\ and\ \bibinfo {author} {\bibfnamefont {D.~J.}\ \bibnamefont
  {Gross}},\ }\bibfield  {title} {\bibinfo {title} {An exact prediction of n=4
  supersymmetric yang–mills theory for string theory},\ }\href
  {https://doi.org/10.1063/1.1372177} {\bibfield  {journal} {\bibinfo
  {journal} {Journal of Mathematical Physics}\ }\textbf {\bibinfo {volume}
  {42}},\ \bibinfo {pages} {2896–2914} (\bibinfo {year} {2001})}\BibitemShut
  {NoStop}%
\end{thebibliography}%

\end{document}